%% file: main.tex
\documentclass[runningheads]{llncs}
\usepackage[T1]{fontenc}
\usepackage[colorlinks]{hyperref}
\usepackage{color}
\usepackage{amsmath}
\usepackage{amssymb}
\usepackage{scalerel}
\usepackage{subcaption}
\usepackage{wrapfig}
\usepackage{tikz}
\usetikzlibrary{calc,arrows,shapes,backgrounds,matrix}
\tikzstyle{stochasticc} = [fill, circle, minimum size=0.1cm, inner sep=0.05cm, outer sep=0cm]
\tikzstyle{stochastics} = [fill, rectangle, minimum size=0.1cm, inner sep=0.05cm, outer sep=0cm]

\input{commands}

\title{Simplicity Lies in the Eye of the Beholder:\texorpdfstring{\\}{ }A Strategic Perspective on Controllers in Reactive Synthesis\thanks{Mickael Randour is an F.R.S.-FNRS Senior Research Associate and member of the TRAIL Institute. His work has been supported by the F.R.S.-FNRS under Grants n° F.4520.18 (MIS ManySynth) and n° T.0188.23 (PDR ControlleRS).}}

\titlerunning{A Strategic Perspective on Controllers in Reactive Synthesis}

\author{Mickael Randour}

\authorrunning{Mickael Randour}

\institute{F.R.S.-FNRS \& UMONS -- Université de Mons, Belgium
\email{mickael.randour@umons.ac.be}}

\begin{document}

\maketitle

\begin{abstract}
In the game-theoretic approach to controller synthesis, we model the interaction between a system to be controlled and its environment as a game between these entities, and we seek an appropriate (e.g., winning or optimal) strategy for the system. This strategy then serves as a formal blueprint for a real-world controller. A common belief is that \textit{simple} (e.g., using limited memory) \textit{strategies are better}: corresponding controllers are easier to conceive and understand, and cheaper to produce and maintain.

This invited contribution focuses on the complexity of strategies in a variety of synthesis contexts. We discuss recent results concerning memory and randomness, and take a brief look at what lies beyond our traditional notions of complexity for strategies.
\end{abstract}

\section{Introduction}
\label{sec:intro}

\paragraph{A story of control.} 
Many different fields --- such as control theory, AI, and formal methods --- are concerned with a common problem: how to control an agent embedded within an uncontrollable environment (possibly involving other agents) in order to achieve a given goal. This challenge was already envisioned by Church in the 1950s~\cite{church1957applications}, and is now known as the \textit{controller synthesis} problem.

My motherland is formal methods, and I have been shaped by the teachings of game theory. It is therefore no surprise that I focus on the \textit{game-theoretic approach to controller synthesis}. The rich history linking Church’s problem, logic, automata theory, and games on graphs is expertly narrated by Thomas in~\cite{DBLP:conf/fossacs/Thomas09}.

\paragraph{The game-theoretic metaphor.} 
The approach can be sketched as follows. Given a \textit{specification} that formalizes what the system should and should not do, along with a formal model of the system, we aim to automatically construct a \textit{controller} that ensures correct behavior, regardless of how the environment behaves. 

In the simplest setting, the controller is viewed as a player, and the uncontrollable environment as its adversary. Their interaction is modeled as a \textit{two-player zero-sum game on a graph}, where vertices represent \textit{states} of the system and environment, and edges represent their possible \textit{actions}. The players take turns moving a pebble along this graph according to \textit{strategies}, generating a sequence of vertices called a \textit{play}, which represents a possible behavior of the system. The controller’s goal is to ensure that the play satisfies the specification, encoded as a \textit{winning objective} (a play is winning if it belongs to the specified set). The adversarial environment’s goal is to prevent this --- we are effectively modeling a worst-case scenario. 

\textit{Establishing winning strategies} (i.e., guaranteeing victory regardless of the environment’s choices) \textit{corresponds to synthesizing implementable models of provably correct controllers}: these strategies are formal blueprints for acceptable controllers~\cite{randourECCS,DBLP:reference/mc/BloemCJ18}.

The state of the art extends far beyond this basic setting. Richer models incorporate, e.g., quantitative payoffs, trade-offs between objectives, or stochastic transitions. In what follows, we use the term ``games'' as an umbrella for: two-player \textit{antagonistic} games as described above; \textit{Markov decision processes} (MDPs, or $1\frac{1}{2}$-player games, where the system faces a fully probabilistic environment); and \textit{stochastic games} (SGs, or $2\frac{1}{2}$-player games, which combine stochastic transitions with an adversarial opponent). A comprehensive introduction to games on graphs, covering classical results and recent developments, can be found in~\cite{gigi}.

\paragraph{Where it all began.}
Every hero has an origin story. So does every scientific endeavor. I began mine around 2010, at a time when games on graphs were already a well-established area of research, particularly for controller synthesis~\cite{DBLP:conf/dagstuhl/2001automata}.

Historically, most game settings were studied considering a single objective for the system, either qualitative (e.g., reachability, B\"uchi, parity) or quantitative (e.g., mean payoff, shortest path, discounted sum). Remarkably, \textit{pure memoryless strategies} --- which we will discuss shortly --- suffice to play optimally (i.e., winning whenever it is possible to win, or optimizing a given payoff function) in two-player games for virtually all classical objectives from the literature, and for both players (see, e.g., Gimbert and Zielonka's characterization~\cite{DBLP:conf/concur/GimbertZ05}). The situation is similar for MDPs and SGs.

But what is a strategy? Mathematically, it is most often viewed as a \textit{function} that maps histories (the sequence of vertices and actions up to the point of decision) to actions (typically the next vertex to visit). In general, there is no limit to how much a strategy can ``remember,'' though infinite-memory strategies are of limited practical interest. Strategies can also rely on randomness in various ways (we will compare them in Section~\ref{sec:randomness}), the most classical one mapping histories to probability distributions over actions (so-called \textit{behavioral} strategies). Pure memoryless strategies are often seen as the \textit{simplest} kind of strategy: they use neither memory, nor randomness. In essence, they are simply functions from vertices to actions. Therefore, their sufficiency in most (single-objective) games can be seen as a blessing. Indeed, in the context of controller synthesis, the general consensus is that \textit{simple strategies are always better} (e.g.,~\cite{DBLP:conf/tacas/DelgrangeKQR20}): corresponding controllers will be easier to conceive, and cheaper to produce and maintain. Simplicity is also strongly correlated with explainability, another desirable feature.

\paragraph{The first steps.}
My early work contributed to a blooming effort to develop \textit{many-sided synthesis} frameworks, which take into account the interplay between different quantitative (or qualitative) aspects and the resulting trade-offs that naturally arise in applications (e.g., decreasing the response time of a system may require additional computing power and energy consumption). I worked in several directions, such as multi-dimension games (e.g.,~\cite{DBLP:journals/acta/ChatterjeeRR14}), Boolean combinations of objectives (e.g.,~\cite{DBLP:conf/fsttcs/0001PR18}), games with heterogeneous objectives (e.g.,~\cite{DBLP:journals/acta/BouyerMRLL18}), combination of hard worst-case constraints with expected value optimization (e.g.,~\cite{DBLP:journals/iandc/BruyereFRR17}), conjunctions of percentile constraints (e.g.,~\cite{DBLP:journals/fmsd/RandourRS17}), and efficient synthesis under simplicity constraints (e.g.,~\cite{DBLP:conf/tacas/DelgrangeKQR20}).

Moving from single-objective games to many-sided ones takes a heavy toll. First, there is no total order on the performance of strategies, which forces us to consider complex \textit{Pareto frontiers} (and corresponding Pareto-optimal strategies) instead of a single optimal strategy (see an example in Section~\ref{sec:randomness}). Second, the \textit{computational complexity} of solving a game or synthesizing a strategy increases rapidly. Finally, and most importantly in the context of this note, \textit{strategies (at least for the system) typically need to be much more complex} to play Pareto-optimally: they might require memory, either small~\cite{DBLP:journals/lmcs/BrihayeDOR20}, huge~\cite{DBLP:journals/acta/ChatterjeeRR14,DBLP:journals/iandc/Chatterjee0RR15,DBLP:journals/iandc/BruyereFRR17}, or even infinite~\cite{DBLP:journals/iandc/VelnerC0HRR15,DBLP:conf/concur/BruyereHRR19}, randomness~\cite{DBLP:journals/acta/ChatterjeeRR14}, or both~\cite{DBLP:journals/fmsd/RandourRS17}. I wrote an introduction to multi-objective games, discussing these costs, in~\cite{chapter_multi}.

\paragraph{Welcome to the Machine.}
The attentive reader may have noticed that, until now, strategies were treated as abstract mathematical objects; yet we now suggest that their (somewhat undefined) memory can be measured. This shift warrants discussion. Many papers on games adopt a purely theoretical view of strategies, often without considering their concrete representation; this is especially common when pure memoryless strategies suffice, as these are frequently viewed as ``trivial to implement.''  
When strategies involving memory are considered, they are typically represented using finite-state machines with outputs --- either \textit{Mealy} or \textit{Moore machines}. From this perspective, the memory of a strategy corresponds to the number of states in its associated machine, and its ``complexity'' is often quantified by this number alone.  
We will return to this point shortly. For an introduction to Mealy machines and their various forms, see the survey~\cite{DBLP:conf/fsttcs/BouyerRV22}.

\paragraph{One meta-theorem to rule them all.}  
Delving into the realm of multi-objective games opens a Pandora's box of endless specific combinations. A flurry of work has explored these combinations --- often seemingly closely related and yielding similar results, yet always differing just enough to leave an independent observer wondering: What is common? What differs? Are similar results obtained for similar reasons? Can we gain transversal insight into multi-objective games and their strategies via a theoretical cross-section?

These were the questions I began to ponder after several years of working on many-sided synthesis. I was particularly inspired by Gimbert and Zielonka's characterization of memoryless-determined two-player games~\cite{DBLP:conf/concur/GimbertZ05}, an elegant result obtained in an elegant way. I was especially drawn to the ``one-to-two-player lift'' established as a corollary: it essentially states that if memoryless strategies suffice in all one-player games (which are much easier to study), for both players, then they also suffice in two-player games.

I started working with Youssouf Oualhadj, aiming to prove a similar result for finite-memory strategies. The project stayed on the back burner for a while but came back to the forefront during Pierre Vandenhove's PhD thesis, which I co-supervised with Patricia Bouyer. In collaboration with Stéphane Le Roux, we were able to prove that such a lift does not hold in full generality. However, we did provide a \textit{comparable characterization and lift for an appropriate class of finite-memory strategies}~\cite{DBLP:journals/lmcs/BouyerRORV22}. This opened new avenues of research for me and triggered a gradual shift toward establishing \textit{meta-theorems} --- aimed at understanding the nature of strategy complexity across diverse settings.

The bulk of this note is dedicated to some of the results obtained in that direction, along the two main dimensions of strategy complexity in the classical model: \textit{memory}~\cite{DBLP:journals/lmcs/BouyerRORV22,DBLP:journals/lmcs/BouyerORV23,DBLP:journals/theoretics/BouyerRV23} and \textit{randomness}~\cite{kuhn_IandC,cocktail}. These results were made possible thanks to wonderful collaborators (see below), and progress along each axis was driven by outstanding PhD students: Pierre Vandenhove for memory, and James C.~A.~Main for randomness.

\paragraph{A rabbit hole.}
Let us take a step back. Simplicity is the paragon virtue of strategies, and we now have a clearer understanding of complexity requirements across a wide range of game models. Is that the end of the story? Not quite.

Indeed, all these results are deeply tied to the \textit{Mealy machine representation} of strategies. This is the predominant model used in the literature (e.g.,~\cite{gigi}) and it is often treated as canonical. Its use is natural, given the central role of logic and automata in synthesis approaches, and it is a versatile tool from a theoretical standpoint. Nonetheless, I will argue in Section~\ref{sec:beyond} that our shared understanding of ``simplicity'' is, in part, a \textit{side-effect of this representational choice} --- one that is somewhat disconnected from the intuitive notion of simplicity one might expect in the context of controller synthesis.

While recent work has begun to explore \textit{alternative representations} --- strategy machines~\cite{DBLP:phd/dnb/Gelderie14}, decision trees~\cite{DBLP:conf/cav/BrazdilCCFK15,DBLP:conf/tacas/BrazdilCKT18}, neural networks~\cite{DBLP:conf/ijcai/CarrJT20}, programs~\cite{gridworlds}, or structurally-enriched Mealy machines~\cite{DBLP:conf/cav/BlahoudekB0OTT20,DBLP:conf/icalp/AjdarowM0R25} --- these efforts remain largely motivated by practical considerations. We are only scratching the surface: a thorough understanding of the \textit{ecosystem of representation model}s --- and their interrelations --- is still lacking, as is a \textit{representation-agnostic theory} of complexity.

\paragraph{Outline.}
A word of warning: given the breadth of models and results covered in this survey, we adopt a somewhat informal approach. There will be occasional hand-waving and approximations to keep the discussion at a high level. Pointers to the relevant papers are provided throughout for full formal details.

In Section~\ref{sec:prelims}, we recall the main concepts of games on graphs, including the Mealy machine representation of strategies. Sections~\ref{sec:memory} and~\ref{sec:randomness} are devoted to results on memory and randomness complexity, respectively, within this model. Section~\ref{sec:beyond} questions our traditional (Mealy-based) notions of complexity and opens a window onto what lies beyond. We conclude briefly in Section~\ref{sec:concl}.

\begin{credits}
\subsubsection{\ackname} 
Most of the results presented in this survey~\cite{DBLP:journals/lmcs/BouyerRORV22,DBLP:journals/lmcs/BouyerORV23,DBLP:journals/theoretics/BouyerRV23,kuhn_IandC,cocktail} were obtained in the context of the F.R.S.-FNRS projects \textit{ManySynth} and \textit{ControlleRS}.

I want to highlight the pivotal role of my outstanding PhD students, James C.~A. Main and Pierre Vandenhove, whose work has been instrumental throughout. I also express my deepest gratitude to my wonderful co-authors: Patricia Bouyer, Stéphane Le Roux, and Youssouf Oualhadj.
\end{credits}

\section{Games for Controller Synthesis}
\label{sec:prelims}
We assume that the reader is familiar with basic concepts in games on graphs. While we recall and illustrate the main notions throughout, we often omit formal definitions. For a comprehensive treatment of the underlying theory, we refer to the recent textbook~\cite{gigi}, which provides all necessary background.

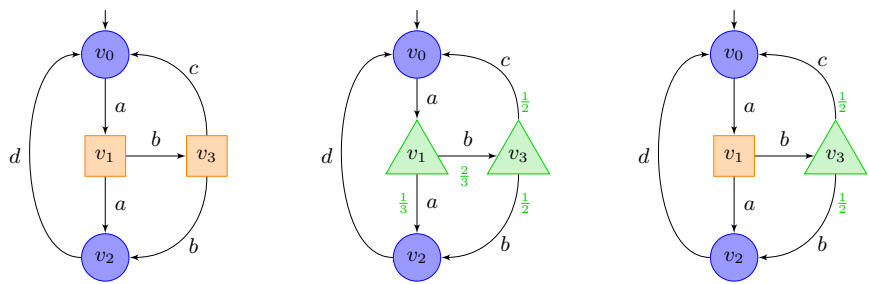
\begin{figure}[tbh]
\centering
\vspace{-5mm}
\begin{subfigure}[t]{0.3\textwidth}
\centering
\scalebox{0.9}{\begin{tikzpicture}[every node/.style={font=\small,inner sep=1pt}]
\draw (0,0) node[rond,bleu] (s0) {$\state_0$};
\draw (0,-1.5) node[carre,orange] (s1) {$\state_1$};
\draw (0,-3) node[rond,bleu] (s2) {$\state_2$};
\draw (1.5,-1.5) node[carre,orange] (s3) {$\state_3$};
\draw[latex'-] (s0.90) -- +(90:3mm);
\draw (s0) edge[-latex'] node[right, xshift=1mm] {$a$} (s1);
\draw (s1) edge[-latex'] node[right, xshift=1mm] {$a$}  (s2);
\draw (s1) edge[-latex'] node[above, yshift=1mm] {$b$} (s3);
\draw (s3) edge[-latex',out=90,in=0,looseness=4,distance=7mm] node[above right,xshift=0mm,yshift=0mm] {$c$} (s0);
\draw (s3) edge[-latex',out=270,in=0,looseness=4,distance=7mm] node[below right,xshift=0mm,yshift=0mm] {$b$} (s2);
\draw (s2) edge[-latex',out=180,in=180,looseness=4,distance=10mm] node[left, xshift=-1mm] {$d$} (s0);
\end{tikzpicture}}
\caption{Two-player game: $\protect \pSys$ and $\pEnv$.}
\label{fig:twoGame}
\end{subfigure}
\hfill
\begin{subfigure}[t]{0.33\textwidth}
\centering
\scalebox{0.9}{\begin{tikzpicture}[every node/.style={font=\small,inner sep=1pt}]
\draw (0,0) node[rond,bleu] (s0) {$\state_0$};
\draw (0,-1.5) node[triang,vert] (s1) {$\state_1$};
\draw (0,-3) node[rond,bleu] (s2) {$\state_2$};
\draw (1.5,-1.5) node[triang,vert] (s3) {$\state_3$};
\draw (-0.2,-2.2) node (l0) {\color{dvert} \scriptsize $\frac{1}{3}$};
\draw (0.7,-1.8) node (l1) {\color{dvert} \scriptsize $\frac{2}{3}$};
\draw (1.6,-0.7) node (l0) {\color{dvert} \scriptsize $\frac{1}{2}$};
\draw (1.6,-2.2) node (l1) {\color{dvert} \scriptsize $\frac{1}{2}$};
\draw[latex'-] (s0.90) -- +(90:3mm);
\draw (s0) edge[-latex'] node[right, xshift=1mm] {$a$} (s1);
\draw (s1) edge[-latex'] node[right, xshift=1mm] {$a$}  (s2);
\draw (s1) edge[-latex'] node[above, yshift=1mm] {$b$} (s3);
\draw (s3) edge[-latex',out=90,in=0,looseness=4,distance=7mm] node[above right,xshift=0mm,yshift=0mm] {$c$} (s0);
\draw (s3) edge[-latex',out=270,in=0,looseness=4,distance=7mm] node[below right,xshift=0mm,yshift=0mm] {$b$} (s2);
\draw (s2) edge[-latex',out=180,in=180,looseness=4,distance=10mm] node[left, xshift=-1mm] {$d$} (s0);
\end{tikzpicture}}
\caption{Markov decision process: $\protect \pSys$ and $\pRand$.}
\label{fig:MDP}
\end{subfigure}
\hfill
\begin{subfigure}[t]{0.3\textwidth}
\centering
\scalebox{0.9}{\begin{tikzpicture}[every node/.style={font=\small,inner sep=1pt}]
\draw (0,0) node[rond,bleu] (s0) {$\state_0$};
\draw (0,-1.5) node[carre,orange] (s1) {$\state_1$};
\draw (0,-3) node[rond,bleu] (s2) {$\state_2$};
\draw (1.5,-1.5) node[triang,vert] (s3) {$\state_3$};
\draw (1.6,-0.7) node (l0) {\color{dvert} \scriptsize $\frac{1}{2}$};
\draw (1.6,-2.2) node (l1) {\color{dvert} \scriptsize $\frac{1}{2}$};
\draw[latex'-] (s0.90) -- +(90:3mm);
\draw (s0) edge[-latex'] node[right, xshift=1mm] {$a$} (s1);
\draw (s1) edge[-latex'] node[right, xshift=1mm] {$a$}  (s2);
\draw (s1) edge[-latex'] node[above, yshift=1mm] {$b$} (s3);
\draw (s3) edge[-latex',out=90,in=0,looseness=4,distance=7mm] node[above right,xshift=0mm,yshift=0mm] {$c$} (s0);
\draw (s3) edge[-latex',out=270,in=0,looseness=4,distance=7mm] node[below right,xshift=0mm,yshift=0mm] {$b$} (s2);
\draw (s2) edge[-latex',out=180,in=180,looseness=4,distance=10mm] node[left, xshift=-1mm] {$d$} (s0);
\end{tikzpicture}}
\caption{Stochastic game: $\protect \pSys$, $\pEnv$, and $\pRand$.}
\label{fig:SG}
\end{subfigure}
\vspace{-2mm}
\caption{Three types of games. Vertices are partitioned between those of $\protect \pSys$, the system, $\protect \pEnv$, an antagonistic adversary, and $\pRand$, a random adversary. Letters are the colors of the edges, whereas fractions represent probability distributions in random vertices. The initial vertex is highlighted via a short arrow.}
\end{figure}

\paragraph{Two-player games.} 
We consider two players, $\pSys$, the controller, and $\pEnv$, its adversarial environment. A two-player turn-based \textit{arena} is a tuple $\arena = (\statesSys, \statesEnv, \edges)$, where $\states = \statesSys \uplus \statesEnv$ is the set of vertices, partitioned into vertices of $\pSys$ and $\pEnv$, and $\edges$ is the set of edges. We assume there are no deadlocks. Each edge $\edge$ connects a source vertex $\In(\edge)$ to a target vertex $\Out(\edge)$. An example is given in Figure~\ref{fig:twoGame}. We assume arenas are finite unless otherwise specified.

We use a \textit{coloring function} $\colFct\colon \edges \to \colors$. Colors may represent, e.g., symbols, priorities (in parity objectives), or weights (in quantitative objectives). They serve as a general abstraction for defining objectives.

A \textit{play} is an infinite sequence $\play = \state_0\edge_1\state_1\edge_2\state_2\ldots{} \in (\states\edges)^\omega$ such that $\state_0$ is a given initial vertex, and for all $i$, $\In(\edge_i) = \state_{i-1}$ and $\Out(\edge_i)=\state_i$ --- we allow multiple edges (with potentially different colors) between two given states; hence, it is necessary to consider edges explicitly. We sometimes identify a play with its projection onto colors, i.e., the sequence $\colFct(\edge_1)\colFct(\edge_2)\ldots{}$. For instance, in Figure~\ref{fig:twoGame}, we would consider $\play = abc(abbd)^\omega$ as a play.

\paragraph{Objectives.} As discussed in Section~\ref{sec:intro}, $\pSys$ aims to enforce a \textit{specification}, while $\pEnv$ attempts to prevent this. The three main ways to formalize a specification using colors are as follows:
\begin{enumerate}
\item A \textit{winning condition} is a set of plays that $\pSys$ aims to realize. E.g., $\mathsf{Reach}(t) = \{ \play = \col_0 \col_1 \col_2 \ldots \mid t \in \pi\}$, for $t \in \colors$ a given color, a \textit{reachability} objective.
\item A \textit{payoff function} represents a quantity to optimize, assuming $\colors \subset \mathbb{Q}$. E.g., the \textit{discounted sum} function, defined as $\mathsf{DS(\pi)} = \sum_{i=0}^{\infty} \gamma^i c_i$ for $\gamma \in (0, 1)$.
\item A \textit{preference relation} defines a total preorder over sequences of colors, thus generalizing both previous concepts.
\end{enumerate}

\paragraph{Strategies and optimality.} Player $\pArb$, with $\arbSign \in \{\raisebox{0.4mm}{\sysSign}, \envSign\}$, selects outgoing edges (from vertices in $\statesArb$) according to a \textit{strategy} $\stratArb \colon (\states\edges)^{\ast} \statesArb \to \edges$, which must be consistent with the underlying graph. Observe that there is no bound on the length of the input history. For now, we consider only strategies that do not involve randomness --- these are called \textit{pure} strategies. We will return to randomized strategies in Section~\ref{sec:randomness}. 

A particularly important subclass is that of \textit{memoryless} strategies, i.e., functions $\stratArb \colon \statesArb \to \edges$, where the choice depends only on the current vertex.

In the context of synthesis, we are interested in the complexity of \textit{optimal} strategies. Recall that we currently focus on single-objective games. Suppose the objective is described by a preference relation $\prefRel$ for $\pSys$. Then a strategy $\stratSys$ is optimal if its worst-case outcome (i.e., considering all strategies of the adversary $\pEnv$) is at least as good, with respect to $\prefRel$, as that of any other strategy $\stratSys'$. That is, we assume a rational adversary playing optimally and evaluate strategies based on the quality of their worst-case outcomes. See~\cite{DBLP:journals/lmcs/BouyerRORV22} for a thorough discussion of optimality under preference relations.

\paragraph{Stochastic environments.} 
The environment might not be fully antagonistic but exhibit stochastic behavior: we introduce two other types of games that incorporate a random player, denoted $\pRand$, representing probabilistic choices.

Given a finite set $S$, we write $\dists(S)$ for the set of probability distributions over $S$. A \textit{Markov decision process (MDP)} is a tuple $\mdp = (\statesSys, \statesRand, \edges, \prTrans)$, where $\states = \statesSys \uplus \statesRand$ is the set of vertices, partitioned into $\pSys$- and $\pRand$-vertices. The edge set is $\edges$, and $\prTrans$ is the probabilistic transition function $\prTrans \colon \statesRand \to \dists(\edges)$, assigning probability distributions to the outgoing edges of random vertices. An example MDP is shown in Figure~\ref{fig:MDP}. While MDPs are often defined using actions (see, e.g.,~\cite{DBLP:journals/fmsd/RandourRS17}), we adopt the \textit{random vertex} formalism for uniformity.

Fixing a strategy $\stratSys$ and an initial vertex $v$ induces a fully stochastic process: a \textit{Markov chain (MC)}. The objective of $\pSys$ can be to maximize either the probability $\prob^{\stratSys}_{v}[\winCond]$ of a winning condition $\winCond$, or the expected value $\expct^{\stratSys}_{v}[\payoff]$, for a payoff function $\payoff$. For the construction of the probability measure, see~\cite{gigi}.

\textit{Stochastic games (SGs)} combine the adversarial and stochastic models of the environment. An SG is a tuple $\sg = (\statesSys, \statesEnv, \statesRand, \edges, \prTrans)$, extending the structures above. An example appears in Figure~\ref{fig:SG}. In SGs, strategies are required for both $\pSys$ and $\pEnv$ to induce an MC. The goal of $\pSys$ is to maximize, against an optimal adversary $\pEnv$ (i.e., worst-case scenario), the quantity $\prob^{\stratSys, \stratEnv}_{v}[\winCond]$ or $\expct^{\stratSys, \stratEnv}_{v}[\payoff]$. 

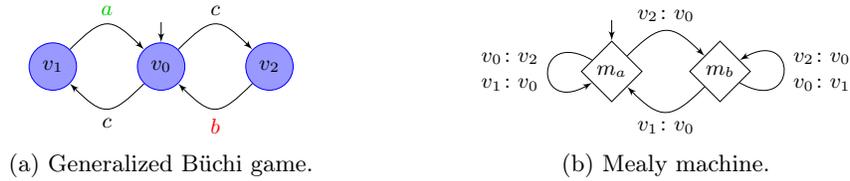
\begin{figure}[tbh]
\centering
\begin{subfigure}[t]{0.4\textwidth}
\centering
\scalebox{0.9}{\begin{tikzpicture}[every node/.style={font=\small,inner sep=1pt}]
\draw (0,0) node[rond,bleu] (s0) {$\state_0$};
\draw (-1.6,0) node[rond,bleu] (s1) {$\state_1$};
\draw (1.6,0) node[rond,bleu] (s2) {$\state_2$};
\draw[latex'-] (s0.90) -- +(90:3mm);
\draw (s0) edge[-latex',out=45,in=135,looseness=4,distance=7mm] node[above,yshift=1mm] {$c$} (s2);
\draw (s1) edge[-latex',out=45,in=135,looseness=4,distance=7mm] node[above,yshift=1mm,dvert] {$a$} (s0);
\draw (s2) edge[-latex',out=225,in=315,looseness=4,distance=7mm] node[below,yshift=-1mm,drouge] {$b$} (s0);
\draw (s0) edge[-latex',out=225,in=315,looseness=4,distance=7mm] node[below,yshift=-1mm] {$c$} (s1);
\end{tikzpicture}}
\caption{Generalized B\"uchi game.}
\label{fig:genBuchi}
\end{subfigure}
\hfill
\begin{subfigure}[t]{0.5\textwidth}
\centering
\scalebox{0.9}{\begin{tikzpicture}[every node/.style={font=\small,inner sep=1pt}]
\draw (0,0) node[diamant] (mb) {$\memState_a$};
\draw (1.6,0) node[diamant] (mc) {$\memState_b$};
\draw[latex'-] (mb.90) -- +(90:3mm);
\draw (mb) edge[-latex',out=45,in=135,looseness=4,distance=7mm] node[above,yshift=1mm] {$\state_2\colon \state_0$} (mc);
\draw (mc) edge[-latex',out=225,in=315,looseness=4,distance=7mm] node[below,yshift=-1mm] {$\state_1\colon \state_0$} (mb);
\draw (mb) edge[-latex',out=150,in=210,looseness=4,distance=1cm] node[left,xshift=-1mm,align=center] {$\state_0\colon \state_2$\\$\state_1\colon \state_0$} (mb);
\draw (mc) edge[-latex',out=330,in=30,looseness=4,distance=1cm] node[right,xshift=1mm,align=center] {$\state_2\colon \state_0$\\$\state_0\colon \state_1$} (mc);
\end{tikzpicture}}
\caption{Mealy machine.}
\label{fig:genBuchiMealy}
\end{subfigure}
\caption{Memory is needed to see {\color{dvert}$a$} and {\color{drouge}$b$} infinitely often. A winning strategy for $\protect \pSys$ is given as a two-state Mealy machine.}
\label{fig:multiObj}
\end{figure}

\paragraph{Multiple objectives.} 
Complex objectives arise when \textit{combining} simple ones, and they usually require more complex strategies to play optimally. Consider the simple (one-player) game depicted in Figure~\ref{fig:genBuchi}: the objective of $\pSys$ is to see both colors {\color{dvert}$a$} and {\color{drouge}$b$} infinitely often. This \textit{generalized B\"uchi objective} requires memory (in this case, a single bit to alternate between $\state_1$ and $\state_2$), whereas single-objective B\"uchi games admit memoryless optimal strategies.

The classical notion of optimality is lost when moving to multi-objective games, as there is usually no total order on plays. Instead, one must consider \textit{Pareto-optimal} strategies and the \textit{Pareto frontier} --- the set of all payoff vectors not dominated by another. We will revisit this in Section~\ref{sec:randomness}.

\paragraph{Mealy machines.}
The classical representation of a strategy $\stratArb$ is via a \textit{Mealy machine}, i.e., a tuple $\mealy = (\memStates, \memStateInit, \nxtFct, \updFct)$, where $\memStates$ is the set of memory states, $\memStateInit$ is the initial one, $\nxtFct\colon \memStates \times \states \to \edges$ is the next-action function, and $\updFct\colon \memStates \times \edges \to \memStates$ is the update function. A Mealy machine defines a finite-memory strategy when $\vert\memStates\vert < \infty$, and a memoryless strategy when $\vert\memStates\vert = 1$.

We illustrate this in Figure~\ref{fig:genBuchiMealy}, where we simplify notation by identifying edges with their destination vertices. The memory state $\memState_a$ (resp.~$\memState_b$) encodes that the last ``interesting'' color observed was {\color{dvert}$a$} (resp.~{\color{drouge}$b$}); we ignore occurrences of $c$, as they are neutral with respect to the objective. Transitions in the diagram represent the combined effect of $\nxtFct$ and $\updFct$. For example, at the first step of the game, the memory is in state $\memState_a$. Since we are in vertex $\state_0$, we follow the only corresponding transition: the next action is chosen to be $\state_2$ and we update the memory to $\memState_a$. Following this strategy, the resulting play is $\play = (c{\color{dvert}a}c{\color{drouge}b})^\omega$: i.e., $\pSys$ will alternate between $\state_1$ and $\state_2$, and satisfy the objective.

\paragraph{The ice-cream conundrum.} Mealy machines are like ice cream: everyone has a favorite flavor and assumes it should be universally regarded as the best and canonical one. The machine above uses \textit{chaotic} memory: its updates depend on the actual edges taken. But many other “flavors” exist: \textit{chromatic} memory (which only depends on the color of edges), memory with or without \textit{$\varepsilon$-transitions}, variants incorporating different forms of \textit{randomness}, and more. We will encounter some of these, but refer the interested reader to~\cite{DBLP:conf/fsttcs/BouyerRV22,gigi} for a broader survey.

Importantly, results about strategy complexity can vary significantly depending on the chosen model --- a nuance that is often overlooked in the literature.

\section{Memory}
\label{sec:memory}

\paragraph{Memoryless strategies.} Our starting point was Gimbert and Zielonka's characterization of \textit{memoryless-determined two-player deterministic games}~\cite{DBLP:conf/concur/GimbertZ05}.

\begin{theorem}[Characterization~\cite{DBLP:conf/concur/GimbertZ05}]
Given a preference relation $\prefRel$, memoryless strategies suffice to play optimally for both players in all finite arenas if and only if $\prefRel$ and its inverse $\prefRel^{-1}$ are monotone and selective.
\end{theorem}

Intuitively, monotony corresponds to stability under prefix addition, while selectivity captures stability under cycle mixing. For example, if $\play$ and $\play'$ are two plays, and we construct a third play $\play''$ by interleaving them, then selectivity ensures that $\play''$ cannot be strictly better than both $\play$ and $\play'$ with respect to $\prefRel$.

Beyond this characterization, Gimbert and Zielonka also derived a corollary of particular interest.

\begin{corollary}[One-to-two-player lift~\cite{DBLP:conf/concur/GimbertZ05}]
If $\prefRel$ is such that (a) in all finite $\pSys$-arenas, $\pSys$ has optimal memoryless strategies, and (b) in all finite $\pEnv$-arenas, $\pEnv$ has optimal memoryless strategies, then both players have optimal memoryless strategies in all finite two-player arenas.
\end{corollary}

One-player games (i.e., graphs) are significantly simpler. The corollary also highlights that no additional memory is needed to deal with an adversary.

\paragraph{Handling finite memory.} While memoryless strategies suffice for most single-objective games, multi-objective ones may require \textit{finite} (as in the generalized B\"uchi case in Section~\ref{sec:prelims}) or \textit{infinite} memory (e.g., for multi-dimension mean-payoff games~\cite{DBLP:journals/iandc/VelnerC0HRR15}). It is natural to ask whether a finite-memory analogue of Gimbert and Zielonka’s result exists. Unfortunately, this hope does not hold: together with Bouyer, Le Roux, Oualhadj, and Vandenhove, we showed that there are objectives for which both players admit finite-memory optimal strategies in all one-player games, yet infinite memory is required in two-player games~\cite{DBLP:journals/lmcs/BouyerRORV22}.

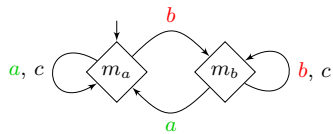
\begin{wrapfigure}{r}{5.4cm}
\centering
\vspace{-6.6mm}
\scalebox{0.9}{\begin{tikzpicture}[every node/.style={font=\small,inner sep=1pt}]
\draw (0,0) node[diamant] (mb) {$m_a$};
\draw (1.6,0) node[diamant] (mc) {$m_b$};
\draw[latex'-] (mb.90) -- +(90:3mm);
\draw (mb) edge[-latex',out=45,in=135,looseness=4,distance=7mm] node[above,yshift=1mm] {{\color{drouge}$b$}} (mc);
\draw (mc) edge[-latex',out=225,in=315,looseness=4,distance=7mm] node[below,yshift=-1mm] {{\color{dvert}$a$}} (mb);
\draw (mb) edge[-latex',out=150,in=210,looseness=4,distance=1cm] node[left,xshift=-1mm,align=center] {{\color{dvert}$a$}, $c$} (mb);
\draw (mc) edge[-latex',out=330,in=30,looseness=4,distance=1cm] node[right,xshift=1mm,align=center] {{\color{drouge}$b$}, $c$} (mc);
\end{tikzpicture}}
\vspace{-3mm}
\caption{A chromatic memory that suffices for $\winCond = \mathsf{Buchi}({\color{dvert}a}) \cap \mathsf{Buchi}({\color{drouge}b})$ in all finite arenas.}
\label{fig:chromatic}
\vspace{-8mm}
\end{wrapfigure}

This led us to define a new frontier: understanding the limits of one-to-two-player lifts. We introduced \textit{arena-independent chromatic memory structures} as suitable tools for this task. Let us revisit the generalized B\"uchi example. Assume $\colors = \{a, b, c\}$ and consider the winning condition $\winCond = \mathsf{Buchi}({\color{dvert}a}) \cap \mathsf{Buchi}({\color{drouge}b})$, which requires both colors to appear infinitely often. Figure~\ref{fig:chromatic} shows such a memory structure: it is \textit{chromatic}, meaning that it depends only on the colors of the edges, not the edges themselves, and thus can be made \textit{arena-independent}. That the structure suffices in all arenas means that it is always possible to define a suitable $\nxtFct$ to build an optimal Mealy machine atop it. In other words, it is possible to play optimally with a memoryless strategy when considering the product arena between the original game and this memory structure.

Using this core concept, we obtained a finite-memory analogue of Gimbert and Zielonka's~\cite{DBLP:journals/lmcs/BouyerRORV22}. Our approach relies on generalizations of monotony and selectivity interpreted \textit{modulo a memory structure}.
\begin{theorem}[Characterization~\cite{DBLP:journals/lmcs/BouyerRORV22}]
Let $\struct$ be a memory structure and $\prefRel$ a preference relation. Strategies based on $\struct$ suffice to play optimally for both players in all finite arenas if and only if $\prefRel$ and $\prefRel^{-1}$ are $\struct$-monotone and $\struct$-selective.
\end{theorem}

\begin{corollary}[One-to-two-player lift~\cite{DBLP:journals/lmcs/BouyerRORV22}]
Let $\struct$ be a memory structure. If $\prefRel$ is such that (a) in all finite $\pSys$-arenas, $\pSys$ has optimal strategies based on $\struct$, and (b) in all finite $\pEnv$-arenas, $\pEnv$ has optimal strategies based on $\struct$, then both players have optimal strategies based on $\struct$ in all finite two-player arenas.
\end{corollary}

\paragraph{Stochastic games.} In joint work with Bouyer, Oualhadj, and Vandenhove, we further extended this result to \textit{pure} arena-independent finite-memory strategies in stochastic games~\cite{DBLP:journals/lmcs/BouyerORV23}, notably establishing a \textit{lift from MDPs to SGs}.

\paragraph{Infinite deterministic arenas.} With Bouyer and Vandenhove, we also investigated arenas of arbitrary cardinality, allowing infinite branching~\cite{DBLP:journals/theoretics/BouyerRV23}. Memory needs can drastically increase: e.g., (one-dimension) mean-payoff objectives require infinite memory whereas memoryless strategies suffice in finite arenas.

It had long been known that all $\omega$-regular winning conditions admit finite-memory optimal strategies in every infinite arena~\cite{Mostowski,ZIELONKA1998135}. We recently established the converse, thereby providing a \textit{complete game-theoretic characterization of $\omega$-regularity}. Chromatic memory structures again played a central role: given a memory structure $\struct$ sufficient for a winning condition $\winCond$, we constructed a parity automaton recognizing $\winCond$, thus proving its $\omega$-regularity. This construction also relies on an auxiliary structure --- the \textit{prefix-classifier} of $\winCond$ --- which distinguishes equivalence classes of histories in the Myhill-Nerode sense.

\begin{theorem}[\cite{DBLP:journals/theoretics/BouyerRV23}]
Chromatic-finite-memory strategies suffice for a winning condition $\winCond$ in all infinite arenas if and only if $\winCond$ is $\omega$-regular. 
\end{theorem}
We also obtain a one-to-two-player lift, now for infinite (deterministic) arenas.

\paragraph{Other criteria and characterizations.} There is a plethora of results related to memory requirements, focusing on a variety of game models. We mention a few here. Casares and Ohlmann provide a characterization through universal graphs~\cite{DBLP:journals/lmcs/CasaresO25}; they notably use the aforementioned $\varepsilon$-transitions. In~\cite{DBLP:conf/lics/CasaresO24}, they also characterize the $\omega$-regular objectives that admit \textit{memoryless} optimal strategies.

Another active line of work focuses on establishing tight memory bounds for particular classes of objectives. We refer in particular to joint work with Bouyer, Casares, Fijalkow, and Vandenhove~\cite{half_pos_LMCS,DBLP:conf/icalp/BouyerFRV23}. While this note has emphasized results that apply to \textit{both} players, some works examine asymmetric requirements: e.g., one player might only require memoryless strategies even if the other one needs (potentially infinite) memory. This situation arises, for instance, in window objective games~\cite{DBLP:journals/iandc/Chatterjee0RR15,DBLP:journals/corr/BruyereHR16}, and is further explored in~\cite{KopThesis,DBLP:journals/amai/BiancoFMM11,DBLP:journals/ijgt/GimbertK23,half_pos_LMCS}.

Other lifting principles have also been studied. Le Roux and Pauly established a \textit{two-to-multi-player lift}~\cite{DBLP:journals/corr/0001P16a}, and in joint work with them we proposed a \textit{one-to-multi-objective lift}~\cite{DBLP:conf/fsttcs/0001PR18}, identifying conditions under which finite-memory determinacy is preserved under combination of objectives.

Additional perspectives can be found in a survey on chromatic memory~\cite{DBLP:conf/fsttcs/BouyerRV22}, co-authored with Bouyer and Vandenhove, and in a recent survey on memory in reachability problems~\cite{DBLP:conf/fsttcs/BrihayeGMR23}, with Brihaye, Goeminne, and Main.

\section{Randomness}
\label{sec:randomness}

\begin{wrapfigure}{r}{40mm}
\centering
\vspace{-6mm}
\scalebox{0.9}{\begin{tikzpicture}[every node/.style={font=\small,inner sep=1pt}]
\draw (0,0) node[rond,bleu] (s0) {$\state_0$};
\draw (-0.6,-1) node[rond,bleu] (s1) {$\state_1$};
\draw (0.6,-1) node[rond,bleu] (s2) {$\state_2$};
\draw[latex'-] (s0.180) -- +(180:3mm);
\draw (s0) edge[-latex'] node[left, yshift=1mm] {$c$} (s1);
\draw (s0) edge[-latex'] node[right, yshift=1mm] {$c$} (s2);
\draw (s1) edge[-latex',out=150,in=210,looseness=4,distance=7mm] node[left,xshift=-1mm,yshift=0mm] {{\color{dvert}$a$}} (s1);
\draw (s2) edge[-latex',out=330,in=30,looseness=4,distance=7mm] node[right,xshift=1mm,yshift=0mm] {{\color{drouge}$b$}} (s2);
\end{tikzpicture}}
\vspace{-2mm}
\caption{Randomness is necessary to see {\color{dvert}$a$} and {\color{drouge}$b$} with non-zero probability.}
\label{fig:coinToss}
\vspace{-8mm}
\end{wrapfigure}
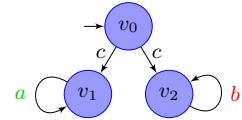

\paragraph{Introducing randomness.} We may need randomness to deal with, e.g., multiple objectives, concurrent games, or imperfect information~\cite{gigi}. Consider the example in Figure~\ref{fig:coinToss}, where the goal of $\pSys$ is to satisfy \(\prob^{\stratSys}[\mathsf{Reach}({\color{dvert}a})] \geq \frac{1}{2} \,\wedge\, \prob^{\stratSys}[\mathsf{Reach}({\color{drouge}b})] \geq \frac{1}{2}.\) Any pure strategy can only lead to two outcomes: either $\state_1$ is \textit{surely} visited, and $\state_2$ never is, or the opposite. Yet, the objective can easily be satisfied by tossing a fair coin in $\state_0$ to pick either $\state_1$ or $\state_2$.

Recall that a \textit{pure} strategy is a function $\stratArb \colon (\states\edges)^{\ast} \statesArb \to \edges$. There are three classical ways to define \textit{randomized} strategies:
\begin{itemize}
\item \textit{behavioral} strategies $\stratArb \colon (\states\edges)^{\ast} \statesArb \to \dists(\edges)$, where a random choice is made \textit{at every decision point};
\item \textit{mixed} strategies $\dists(\stratArb \colon (\states\edges)^{\ast} \statesArb \to \edges)$, where a random choice is made \textit{once at the beginning} of the game to pick a pure strategy to follow; and
\item \textit{general} strategies $\dists(\stratArb \colon (\states\edges)^{\ast} \statesArb \to \dists(\edges))$, which mix both types of randomness.
\end{itemize}
A celebrated result, known as Kuhn's theorem, states that behavioral and mixed strategies are equivalent in games with \textit{perfect recall} --- a property satisfied by all games considered in this note. This result was first established by Kuhn for games on finite trees, and generalized by Aumann~\cite{Aumann64} for infinite games. This equivalence was later extended to general strategies by Bertrand, Genest, and Gimbert~\cite{DBLP:journals/jacm/BertrandGG17}. However, these equivalences rely on \textit{infinite memory} and \textit{probability distributions with infinite support}, which may not be realistic in the context of controller synthesis.
  
\paragraph{Finite-memory strategies.} Based on the Mealy machine representation $\mealy = (\memStates, \memStateInit, \nxtFct, \updFct)$, randomization can be implemented in different ways: via an initial distribution $\distInit \in \dists(\memStates)$, replacing $\memStateInit$;
via the next-action function $\nxtFct\colon \memStates \times \states \to \dists(\edges)$; or  via the update function $\updFct\colon \memStates \times \edges \to \dists(\memStates)$. We classify the resulting strategies using a three-letter code XYZ, where each of X, Y, Z $\in \{\text{D}, \text{R}\}$ denotes whether the initialization, next-action, and update functions are deterministic (D) or randomized (R). For example, DRD refers to strategies with deterministic initialization and update, but randomized action selection --- the natural finite-memory analogue to \textit{behavioral} strategies.

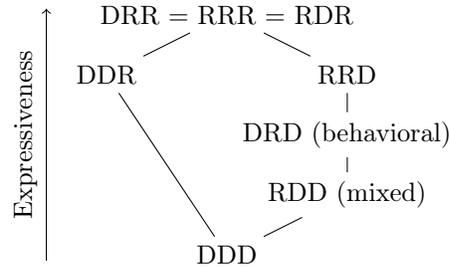
\begin{wrapfigure}{r}{62mm}
\centering
\begin{tikzpicture}
\node at (0, 0) (RRR) {DRR $=$ RRR $=$ RDR};
\node at (-1.6, -0.8) (DDR) {DDR};
\node at (1.6, -0.8) (RRD) {RRD};
\node at (1.6, -1.6) (DRD) {DRD (behavioral)};
\node at (1.6, -2.4) (RDD) {RDD (mixed)};
\node at (0, -3.2) (DDD) {DDD};
\node at (-2.4, 0.2) (top) {};
\node at (-2.4, -3.4) (bot) {};
\draw[->] (bot) -- (top) node[rotate=90,midway,above] {Expressiveness};
\draw (RRR) -- (DDR);
\draw (RRR) -- (RRD);
\draw (RRD) -- (DRD);
\draw (DRD) -- (RDD);
\draw (DDR) -- (DDD);
\draw (RDD) -- (DDD);
\end{tikzpicture}
\vspace{-2mm}
\caption{Taxonomy of randomized finite-memory strategy classes.}
\label{fig:taxonomy}
\vspace{-8mm}
\end{wrapfigure}

In joint work with Main~\cite{kuhn_IandC}, we established a \textit{complete taxonomy} of these strategy classes, illustrated in Figure~\ref{fig:taxonomy}. Each edge indicates a strict increase in expressive power under the notion of \emph{outcome-equivalence}, i.e., the ability to induce the same probability distributions over plays. We directly see that Kuhn’s theorem crumbles under finite memory.

Our taxonomy holds from one-player deterministic games (no collapse) up to concurrent partial-information multi-player games (equivalences still hold). While we consider here a \textit{specification-agnostic equivalence relation}, collapses may arise for restricted classes of objectives (if the additional expressive power is not needed).

\begin{wrapfigure}{r}{50mm}
\centering
\vspace{-6mm}
\scalebox{0.9}{\begin{tikzpicture}[every node/.style={font=\small,inner sep=1pt}]
\draw (0,0) node[rond,bleu] (home) {$h$};
\draw (2.6,0) node[rond,bleu] (work) {$w$};
\draw (1.3,1) node[triang,vert] (train) {$t$};
\draw (1.3,-1) node[triang,vert] (bike) {$b$};
\draw (1.5,-0.6) node (l0) {\scriptsize $1$};
\draw (1.5,0.6) node (l1) {\scriptsize $\frac{1}{4}$};
\draw (1.1,1.4) node (l2) {\scriptsize $\frac{3}{4}$};
\draw[latex'-] (home.180) -- +(180:3mm);
\draw (home) edge[-latex'] node[below, yshift=-1mm] {$0$} (bike);
\draw (home) edge[-latex'] node[above, yshift=1mm] {$0$} (train);
\draw (train) edge[-latex'] node[above, xshift=1mm, yshift=1mm] {$10$} (work);
\draw (train) edge[-latex',out=150,in=90,looseness=4,distance=7mm] node[left,xshift=-2mm,yshift=0mm] {$5$} (home);
\draw (bike) edge[-latex'] node[below, xshift=1mm, yshift=-1mm] {$30$} (work);
\draw (work) edge[-latex',out=330,in=30,looseness=4,distance=7mm] node[right,xshift=1mm,yshift=0mm] {$0$} (work);
\end{tikzpicture}}
\vspace{-0.6mm}
\caption{From \textit{home}, take the \textit{train} or \textit{bike} to reach \textit{work}.}
\label{fig:workMDP}
\vspace{-8mm}
\end{wrapfigure}
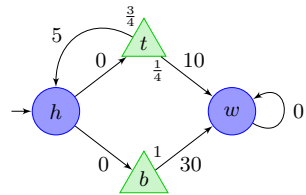

\paragraph{Payoff sets.} Consider the multi-objective MDP in Figure~\ref{fig:workMDP}, modeling an everyday dilemma: $\pSys$ starts at home ($h$) and aims to reach work ($w$). The random vertices model the available modes of transportation: the train ($t$), which is delayed with probability $\frac{3}{4}$, and the bike ($b$), which is deterministic. The time required for each transition is indicated on the edges. The objective is twofold: $(i)$ reaching work under 40 minutes with high probability, and $(ii)$ minimizing the expected time to work. Balancing these two aspects calls for reasoning about \textit{Pareto-optimal strategies}, or more generally, characterizing the full \textit{set of achievable payoffs}.

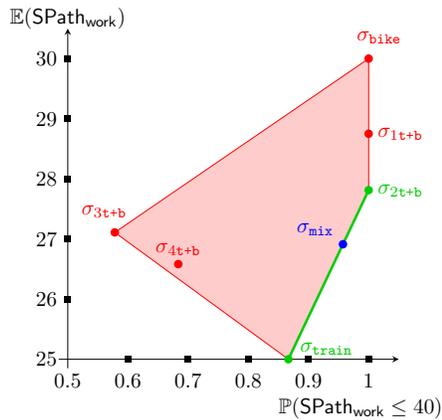
\begin{wrapfigure}{l}{60mm}
\centering
\vspace{-6mm}
\scalebox{0.8}{\begin{tikzpicture}
\draw[-stealth] (-0.15,0) -- (5.5,0) node[yshift=-23,xshift=-18] {$\prob(\mathsf{SPath}_{\textsf{work}} \leq 40)$}; 
\draw[-stealth] (0,-0.15) -- (0,5.5) node[yshift=5] {$\expct(\mathsf{SPath}_{\textsf{work}})$};
\coordinate (oo) at (0, 0);
\coordinate (1x) at (1,0);
\coordinate (2x) at (2,0);
\coordinate (3x) at (3,0);
\coordinate (4x) at (4,0);
\coordinate (5x) at (5,0);
\coordinate (1y) at (0,1);
\coordinate (2y) at (0,2);
\coordinate (3y) at (0,3);
\coordinate (4y) at (0,4);
\coordinate (5y) at (0,5);
\node[xshift=-10] at (oo) {$25$};
\node[stochastics] at (1y) (qy){};
\node[xshift=-10] at (1y) {$26$};
\node[stochastics] at (2y) (qy){};
\node[xshift=-10] at (2y) {$27$};
\node[stochastics] at (3y) (qy){};
\node[xshift=-10] at (3y) {$28$};
\node[stochastics] at (4y) (qy){};
\node[xshift=-10] at (4y) {$29$};
\node[stochastics] at (5y) (qy){};
\node[xshift=-10] at (5y) {$30$};
\node[yshift=-10] at (oo) {$0.5$};
\node[stochastics] at (1x) (qx){};
\node[yshift=-10] at (1x) {$0.6$};
\node[stochastics] at (2x) (qx){};
\node[yshift=-10] at (2x) {$0.7$};
\node[stochastics] at (3x) (qx){};
\node[yshift=-10] at (3x) {$0.8$};
\node[stochastics] at (4x) (qx){};
\node[yshift=-10] at (4x) {$0.9$};
\node[stochastics] at (5x) (qx){};
\node[yshift=-10] at (5x) {$1$};
\coordinate (v0) at (3.6651611328125, 0.0);
\coordinate (v1) at (5.0, 2.8125);
\coordinate (v2) at (5.0, 5);
\coordinate (v3) at (0.78125, 2.109375);
\coordinate (v4) at (3.6651611328125, 0.0);
\coordinate (p1) at (5.0, 3.75);
\coordinate (p4) at (1.8359375, 1.58203125);
\coordinate (p5) at (2.626953125, 1.1865234375);
\coordinate (p6) at (3.22021484375, 0.889892578125);
\coordinate (p7) at (3.6651611328125, 0.66741943359375);
\coordinate (p8) at (3.6651611328125, 0.5005645751953125);
\coordinate (p9) at (3.6651611328125, 0.3754234313964844);
\coordinate (mix) at (4.5728515625, 1.9124999999999979);
\fill[drouge!20] (v0) --(v1) -- (v2) -- (v3) -- (v0);
\draw[drouge] (v0) --(v1) -- (v2) -- (v3) -- (v0);
\draw[dvert,very thick] (v1) -- (v0);
{\color{drouge}
\node[stochasticc,dvert] at (v0) {};
\node[xshift=18, yshift=5, dvert] at (v0) {$\strat_{\texttt{train}}$};
\node[stochasticc] at (v2) {};
\node[yshift=10,xshift=5] at (v2) {$\strat_{\texttt{bike}}$};
\node[stochasticc] at (p1) {};
\node[xshift=15] at (p1) {$\strat_{1\texttt{t+b}}$};
\node[stochasticc,dvert] at (v1) {};
\node[xshift=15,dvert] at (v1) {$\strat_{2\texttt{t+b}}$};
\node[stochasticc] at (v3) {};
\node[yshift=10,xshift=-5] at (v3) {$\strat_{3\texttt{t+b}}$};
\node[stochasticc] at (p4) {};
\node[yshift=7] at (p4) {$\strat_{4\texttt{t+b}}$};}
{\color{dbleu}
\node[stochasticc] at (mix) {};
\node[xshift=-13, yshift=7] at (mix) {$\strat_{\texttt{mix}}$};}
\end{tikzpicture}}
\vspace{-6mm}
\caption{Achievable payoff set.}
\label{fig:payoffSet}
\vspace{-6mm}
\end{wrapfigure}

We depict this set in Figure~\ref{fig:payoffSet}, using the shortest-path payoff function~\cite{DBLP:conf/vmcai/RandourRS15}. A payoff is \textit{Pareto-optimal} if it is not dominated by any other payoff; in our setting, this means that no other point lies in its south-east quadrant. The \textit{Pareto frontier} is thus the line segment from $\strat_{\texttt{train}}$ to $\strat_{2\texttt{t+b}}$, shown in green in the figure.

We highlight strategies of particular interest: $\strat_{\texttt{train}}$ (resp.~$\strat_{\texttt{bike}}$) is the pure memoryless strategy that always picks the train (resp.~the bike). Strategies $\strat_{n\texttt{t+b}}$ try to take the train $n$ times before switching to the bike; they are pure but use memory. Finally, $\strat_{\texttt{mix}}$ represents a \textit{randomized} strategy that \textit{mixes} $\strat_{2\texttt{t+b}}$ and $\strat_{\texttt{train}}$ to achieve a payoff that lies outside the reach of any pure strategy. In this particular example, all achievable payoffs can be obtained via mixed strategies --- the lowest level of the taxonomy above.

Together with Main~\cite{cocktail}, we studied the \textit{structure of payoff sets} in multi-objective MDPs --- they are not necessarily simple polytopes as above. We notably proved that for all payoff functions with finite expectation, the set of achievable payoffs coincides with the convex hull of pure payoffs: from a strategic standpoint, this implies that \textit{mixing a bounded number of pure strategies is sufficient to obtain any achievable payoff} (as in our example). Even when expectations are well-defined but potentially infinite, mixed strategies can still approximate any target payoff arbitrarily closely.

\paragraph{Trading memory for randomness.} We close this section by building a bridge between memory and randomness. Recall the generalized B\"uchi game in Figure~\ref{fig:multiObj}: we previously saw that \textit{pure} strategies require memory to win. Observe that a \textit{behavioral} randomized \textit{memoryless} strategy suffices to win with probability one: playing both $\state_1$ and $\state_2$ with non-zero probability ensures it. This illustrates how, in certain classes of games and objectives, randomness can compensate for the absence of memory. We notably explored this phenomenon for combinations of energy, mean-payoff and parity objectives, in a joint work with Chatterjee and Raskin~\cite{DBLP:journals/acta/ChatterjeeRR14}. Similar investigations can be found in~\cite{DBLP:conf/qest/ChatterjeeAH04,DBLP:conf/hybrid/ChatterjeeHP08,DBLP:conf/stacs/Horn09,DBLP:journals/lmcs/MonmegePR25a}.

\paragraph{Related questions.} While memory requirements are widely studied, much less is known about randomness. Cristau, David and Horn gave a first comparison of the three classical types of randomized strategies in~\cite{DBLP:journals/corr/abs-1006-1404}. Chatterjee et al.~also studied when randomness is useful in strategies or games~\cite{DBLP:journals/iandc/Chatterjee0GH15}.

\section{Beyond Mealy Machines}
\label{sec:beyond}

\paragraph{An incomplete story.} The leitmotiv in controller synthesis is that \textit{simpler strategies are better}.  But what does ``simple'' really mean? The prevailing answer is: small memory and no randomness (e.g.,~\cite{DBLP:conf/tacas/DelgrangeKQR20}) --- with the underlying assumption that strategies \textit{are} Mealy machines. We close by questioning this statement.

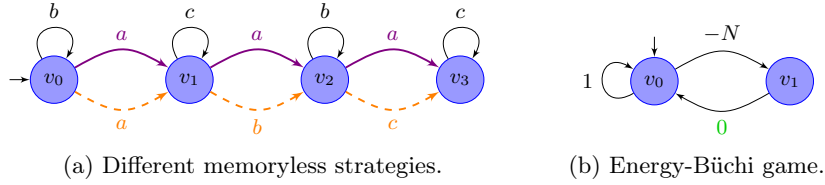
\begin{figure}[tbh]
\centering
\begin{subfigure}[t]{0.6\textwidth}
\centering
\scalebox{0.9}{\begin{tikzpicture}[every node/.style={font=\small,inner sep=1pt}]
\draw (0,0) node[rond,bleu] (s0) {$\state_0$};
\draw (2,0) node[rond,bleu] (s1) {$\state_1$};
\draw (4,0) node[rond,bleu] (s2) {$\state_2$};
\draw (6,0) node[rond,bleu] (s3) {$\state_3$};
\draw[latex'-] (s0.180) -- +(180:3mm);
\draw (s0) edge[-latex',out=120,in=60,looseness=4,distance=7mm] node[above,yshift=1mm] {$b$} (s0);
\draw (s1) edge[-latex',out=120,in=60,looseness=4,distance=7mm] node[above,yshift=1mm] {$c$} (s1);
\draw (s2) edge[-latex',out=120,in=60,looseness=4,distance=7mm] node[above,yshift=1mm] {$b$} (s2);
\draw (s3) edge[-latex',out=120,in=60,looseness=4,distance=7mm] node[above,yshift=1mm] {$c$} (s3);
\draw (s0) edge[-latex',out=30,in=150,looseness=4,distance=7mm,draw=dviolet,thick] node[above,yshift=1mm,dviolet] {$a$} (s1);
\draw (s1) edge[-latex',out=30,in=150,looseness=4,distance=7mm,draw=dviolet,thick] node[above,yshift=1mm,dviolet] {$a$} (s2);
\draw (s2) edge[-latex',out=30,in=150,looseness=4,distance=7mm,draw=dviolet,thick] node[above,yshift=1mm,dviolet] {$a$} (s3);
\draw (s0) edge[-latex',out=330,in=210,looseness=4,distance=7mm,draw=dorange,thick,dashed] node[below,yshift=-1mm,dorange] {$a$} (s1);
\draw (s1) edge[-latex',out=330,in=210,looseness=4,distance=7mm,draw=dorange,thick,dashed] node[below,yshift=-1mm,dorange] {$b$} (s2);
\draw (s2) edge[-latex',out=330,in=210,looseness=4,distance=7mm,draw=dorange,thick,dashed] node[below,yshift=-1mm,dorange] {$c$} (s3);
\end{tikzpicture}}
\caption{Different memoryless strategies.}
\label{fig:memorylessAreNotEqual}
\end{subfigure}
\begin{subfigure}[t]{0.35\textwidth}
\centering
\scalebox{0.9}{\begin{tikzpicture}[every node/.style={font=\small,inner sep=1pt}]
\draw (0,0) node[rond,bleu] (s0) {$\state_0$};
\draw (2,0) node[rond,bleu] (s1) {$\state_1$};
\draw[latex'-] (s0.90) -- +(90:3mm);
\draw (s0) edge[-latex',out=210,in=150,looseness=4,distance=7mm] node[left,xshift=-1mm] {$1$} (s0);
\draw (s0) edge[-latex',out=30,in=150,looseness=4,distance=7mm] node[above,yshift=1mm] {$-N$} (s1);
\draw (s1) edge[-latex',out=210,in=330,looseness=4,distance=7mm] node[below,yshift=-1mm] {{\color{dvert}$0$}} (s0);
\end{tikzpicture}}
\caption{Energy-B\"uchi game.}
\label{fig:exponentialIsOK}
\end{subfigure}
\caption{Strategy complexity is representation-dependent.}
\label{fig:mottoIsBroken}
\vspace{-4mm}
\end{figure}

\paragraph{The limits of our motto.} Not all memoryless strategies are created equal. Consider Figure~\ref{fig:memorylessAreNotEqual} and let us interpret colors as actions: intuitively, strategy {\color{dviolet}$\strat_1$} (solid violet transitions) seems simpler than strategy {\color{dorange}$\strat_2$} (dashed orange transitions) --- it would be easier to explain, verify, or implement. Yet, both are encoded as trivial Mealy machines with a single memory state and are thus treated as equally simple under the standard theoretical framework. This highlights a blind spot: the \textit{representation of the next-action function} is typically reduced to a lookup table, with little consideration for its internal structure or interpretability. In large state spaces, even such memoryless strategies may become unwieldy.

Conversely, quantitative games often require strategies with (at least) exponential memory --- typically seen as prohibitively complex. Yet this is not always a barrier in practice. Consider the game in Figure~\ref{fig:exponentialIsOK}, where the objective combines an energy condition (ensuring a non-negative running sum) and a B\"uchi condition on the {\color{dvert}$0$}-edge. Winning requires cycling in $\state_0$ for $N$ steps before transitioning to $\state_1$, and repeating. Encoding this strategy as a Mealy machine demands $N+1$ distinct memory states --- one per counter value --- making it pseudo-polynomial in size. But from an engineering perspective, such a strategy is easily implementable with a simple counter: the \textit{lack of data structures} in the Mealy machine model is what is inflating the apparent complexity.

These examples suggest that our classical measure of strategy complexity, based on Mealy machines, is deeply \textit{model-dependent}. While theoretically convenient, it may misrepresent practical simplicity. There is value in exploring \textit{alternative representations of strategies} that reflect the true cost of implementation and allow for richer notions of simplicity.

\paragraph{Alternative representations.} We briefly survey several recent alternatives. Most of these approaches are motivated by practical concerns; their theoretical underpinnings remain relatively underdeveloped compared to the rich literature on Mealy machines.

A natural first idea is to augment Mealy machines with data structures to preserve internal structure rather than ``flatten'' it. These \textit{structurally-enriched Mealy machines} have notably been explored for counter-based strategies: see Blahoudek et al.~\cite{DBLP:conf/cav/BlahoudekB0OTT20}, or our recent work with Ajdar\'ow, Main and Novotn\'y~\cite{DBLP:conf/icalp/AjdarowM0R25}. Practically, such models yield \textit{more succinct} and \textit{more explainable} strategies, thanks to their transparent internal logic. They also shine a new light on a theoretical level, changing our view of which strategies are complex or not.

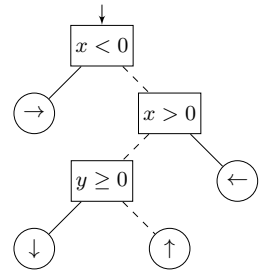
\begin{wrapfigure}{r}{40mm}
\centering
\vspace{-6mm}
\scalebox{0.90}{
\begin{tikzpicture}[every node/.style={font=\small,inner sep=2pt}]
\draw (0,0) node[carre] (t0) {$x < 0$};
\draw (-1,-1) node[roundrect] (a0) {$\rightarrow$};
\draw (1,-1) node[carre] (t1) {$x > 0$};
\draw (2,-2) node[roundrect] (a1) {$\leftarrow$};
\draw (0,-2) node[carre] (t2) {$y \geq 0$};
\draw (1,-3) node[roundrect] (a2) {$\uparrow$};
\draw (-1,-3) node[roundrect] (a3) {$\downarrow$};
\draw[latex'-] (t0.90) -- +(90:3mm);
\draw (t0) edge[-,out=315,in=135,dashed] (t1);
\draw (t0) edge[-,out=225,in=45] (a0);
\draw (t1) edge[-,out=225,in=45,dashed] (t2);
\draw (t1) edge[-,out=315,in=135] (a1);
\draw (t2) edge[-,out=225,in=45] (a3);
\draw (t2) edge[-,out=315,in=135,dashed] (a2);
\end{tikzpicture}}
\caption{Decision tree to reach $(0, 0)$ in a 2D grid.}
\label{fig:DT}
\vspace{-8mm}
\end{wrapfigure}

A well-studied alternative is the use of \textit{decision trees (DTs)}, particularly suited to highly structured state and action spaces~\cite{DBLP:conf/cav/BrazdilCCFK15,DBLP:conf/tacas/BrazdilCKT18}. DTs have primarily been used to compactly represent \textit{memoryless} strategies, serving as a compressed substitute for \textit{large next-action tables}. For instance, consider navigation on a 2D grid with integer coordinates --- say a square $[-d, d] \times [-d, d]$ --- with the objective of reaching the origin. A naïve lookup table would require $\mathcal{O}(d^2)$ entries, while a simple DT (such as that in Figure~\ref{fig:DT}) can represent the strategy compactly for \textit{any} grid size --- even infinite. One key design challenge is balancing the complexity of tests at internal nodes (which affects interpretability) against the overall size of the tree.

Another family of models are \textit{strategy machines}, inspired by Turing machines: they are powerful, yet difficult to synthesize and analyze~\cite{DBLP:phd/dnb/Gelderie14}. Interestingly, this model has led to a tentative notion of \textit{decision speed} --- a rarely explored but potentially meaningful complexity measure. \textit{Neural networks} have become prevalent in reinforcement learning and are increasingly combined with finite-state machines to aid synthesis~\cite{DBLP:conf/ijcai/CarrJT20}; their performance is strong, but interpretability and verifiability remain major obstacles. Finally, \textit{programmatic representations} are closer to realistic code, thus inherently understandable. They are however strongly linked to the input format of the synthesis problem (e.g.,~PRISM code) and hard to generalize~\cite{gridworlds}.

\section{Conclusion}
\label{sec:concl}

In controller synthesis, \textit{simplicity is the name of the game}. But what does it truly mean? We provided a high-level overview of the two classical axes of complexity under the Mealy machine model: \textit{memory} and \textit{randomness}. We then pointed out critical blind spots in this framework, showing that \textit{strategy complexity is often distinct from representation complexity}.

Looking ahead, we argue for the development of a \textit{toolbox of diverse strategy representations}, grounded in a deeper understanding of their expressiveness, succinctness, and interpretability. Ultimately, our goal should be a principled, \textit{representation-agnostic theory of strategy complexity}, capable of guiding both theory and practice in synthesis.

\bibliographystyle{splncs04}
\bibliography{biblio}
\end{document}

%% file: commands.tex
\colorlet{drouge}{red}
\colorlet{frouge}{red!20!white}
\colorlet{dbleu}{blue}
\colorlet{fbleu}{blue!40!white}
\colorlet{dviolet}{blue!50!red}
\colorlet{fviolet}{blue!50!red!40!white}
\colorlet{dvert}{green!80!black}
\colorlet{fvert}{green!80!black!20!white}
\colorlet{djaune}{yellow!80!black}
\colorlet{fjaune}{yellow!80!black!20!white}
\colorlet{dgris}{white!60!black}
\colorlet{fgris}{white!90!black}
\colorlet{dgrisf}{white!30!black}
\colorlet{fgrisf}{white!70!black}
\colorlet{dorange}{red!50!yellow}
\colorlet{forange}{red!50!yellow!30!white}

\tikzstyle{ptrond}=[draw,circle,minimum height=2mm]
\tikzstyle{ptcarre}=[draw,minimum width=3mm,minimum height=3mm]
\tikzstyle{moyrond}=[draw,circle,minimum height=5mm]
\tikzstyle{moycarre}=[draw,minimum width=4mm,minimum height=4mm]
\tikzstyle{rond}=[draw,circle,minimum height=7mm]
\tikzstyle{carre}=[draw,minimum width=6mm,minimum height=6mm]
\tikzstyle{rouge}=[draw=drouge,fill=frouge]
\tikzstyle{vert}=[draw=dvert,fill=fvert]
\tikzstyle{jaune}=[draw=djaune,fill=fjaune]
\tikzstyle{bleu}=[draw=dbleu,fill=fbleu]
\tikzstyle{violet}=[draw=dviolet,fill=fviolet]
\tikzstyle{orange}=[draw=dorange,fill=forange]
\tikzstyle{gris}=[draw=dgris,fill=fgris]
\tikzstyle{grisf}=[draw=dgrisf,fill=fgrisf]
\tikzstyle{rvert}=[style=rond,style=vert]
\tikzstyle{rrouge}=[style=rond,style=rouge]
\tikzstyle{roundrect}=[draw,rounded rectangle, minimum width=6mm,minimum height=6mm]
\tikzstyle{splitrond}=[draw,circle split,minimum height=7mm,circle split part fill={blue!50,red!50}]
\tikzstyle{splitrondbv}=[draw,circle split,minimum height=7mm,circle split part fill={fbleu,fvert}]
\tikzstyle{splitrondrg}=[draw,circle split,minimum height=7mm,circle split part fill={frouge,fgris}]
\tikzstyle{splitrondbo}=[draw,circle split,minimum height=7mm,circle split part fill={fbleu,forange}]
\tikzstyle{accepting}=[edge node={node[scale=1.5] {$\bullet$}}]
\tikzstyle{rond}=[draw,circle,minimum height=7mm]
\tikzstyle{oval}=[draw,ellipse,minimum height=7mm]
\tikzstyle{diamant}=[draw,diamond,minimum height=9mm,minimum width=9mm,aspect=1]
\tikzstyle{carre}=[draw,minimum width=6mm,minimum height=6mm]
\tikzstyle{triang}=[draw,regular polygon,regular polygon sides=3,minimum width=4mm,minimum height=4mm]

\newcommand{\midcirc}{\scaleobj{0.8}{\bigcirc}}
\newcommand{\sysSign}{\ensuremath{\midcirc}}
\newcommand{\envSign}{\ensuremath{\scaleobj{0.9}{\square}}}
\newcommand{\randSign}{\ensuremath{\triangle}}
\newcommand{\arbSign}{\ensuremath{\nabla}}
\newcommand{\pSys}{\ensuremath{\mathcal{P}_{\sysSign}}}
\newcommand{\pEnv}{\ensuremath{\mathcal{P}_{\envSign}}}
\newcommand{\pRand}{\ensuremath{\mathcal{P}_{\randSign}}}
\newcommand{\pArb}{\ensuremath{\mathcal{P}_{\arbSign}}}
\newcommand{\col}{\ensuremath{c}}
\newcommand{\colors}{\ensuremath{C}}
\newcommand{\colFct}{\ensuremath{\mathfrak{c}}}
\newcommand{\arena}{\ensuremath{\mathcal{A}}}
\newcommand{\state}{\ensuremath{v}}
\newcommand{\states}{\ensuremath{V}}
\newcommand{\statesSys}{\ensuremath{V_{\sysSign}}}
\newcommand{\statesEnv}{\ensuremath{V_{\envSign}}}
\newcommand{\statesArb}{\ensuremath{V_{\arbSign}}}
\newcommand{\statesRand}{\ensuremath{V_{\randSign}}}
\newcommand{\edges}{\ensuremath{E}}
\newcommand{\edge}{\ensuremath{e}}
\newcommand{\play}{\ensuremath{\pi}}
\newcommand{\In}{\ensuremath{\mathsf{In}}}
\newcommand{\Out}{\ensuremath{\mathsf{Out}}}
\newcommand{\strat}{\ensuremath{\sigma}}
\newcommand{\stratSys}{\ensuremath{\strat_{\sysSign}}}
\newcommand{\stratEnv}{\ensuremath{\strat_{\envSign}}}
\newcommand{\stratArb}{\ensuremath{\strat_{\arbSign}}}
\newcommand{\prefRel}{\ensuremath{\sqsubseteq}}
\newcommand{\mdp}{\ensuremath{\mathcal{D}}}
\newcommand{\sg}{\ensuremath{\mathcal{S}}}
\newcommand{\prTrans}{\ensuremath{\delta}}
\newcommand{\dists}{\ensuremath{\mathfrak{D}}}
\newcommand{\prob}{\ensuremath{\mathbb{P}}}
\newcommand{\expct}{\ensuremath{\mathbb{E}}}
\newcommand{\winCond}{\ensuremath{W}}
\newcommand{\payoff}{\ensuremath{f}}
\newcommand{\memState}{\ensuremath{m}}
\newcommand{\memStateInit}{\ensuremath{m_{\mathsf{init}}}}
\newcommand{\memStates}{\ensuremath{M}}
\newcommand{\distInit}{\ensuremath{\mu_{\mathsf{init}}}}
\newcommand{\mealy}{\ensuremath{\mathcal{M}}}
\newcommand{\updFct}{\ensuremath{\alpha_{\mathsf{up}}}}
\newcommand{\nxtFct}{\ensuremath{\alpha_{\mathsf{nxt}}}}
\newcommand{\struct}{\ensuremath{\mathcal{M}_{\mathsf{s}}}}